\documentclass[aps,prb,superscriptaddress,twocolumn,amsmath,amssymb,letter]{revtex4}

\usepackage{graphicx}
\usepackage{dcolumn}
\usepackage{bm}

\begin{document}

\title{The hierarchy of multiple many-body interaction scales in high-temperature superconductors}
\author{W. Meevasana}
\email[]{non@stanford.edu} \affiliation {Department of Physics,
Applied Physics, and Stanford Synchrotron Radiation Laboratory,
Stanford University, Stanford, CA 94305}

\author{X.J. Zhou}
\affiliation {Department of Physics, Applied Physics, and Stanford
Synchrotron Radiation Laboratory, Stanford University, Stanford,
CA 94305} \affiliation {Advanced Light Source, Lawrence Berkeley
National Lab, Berkeley, CA 94720}

\author{S. Sahrakorpi}
\affiliation {Physics Department, Northeastern University, Boston,
MA 02115}

\author{W.S. Lee}
\affiliation {Department of Physics, Applied Physics, and Stanford
Synchrotron Radiation Laboratory, Stanford University, Stanford,
CA 94305}

\author{W.L. Yang}
\affiliation {Department of Physics, Applied Physics, and Stanford
Synchrotron Radiation Laboratory, Stanford University, Stanford,
CA 94305} \affiliation {Advanced Light Source, Lawrence Berkeley
National Lab, Berkeley, CA 94720}

\author{K. Tanaka}
\affiliation {Department of Physics, Applied Physics, and Stanford
Synchrotron Radiation Laboratory, Stanford University, Stanford,
CA 94305} \affiliation {Advanced Light Source, Lawrence Berkeley
National Lab, Berkeley, CA 94720}

\author{N. Mannella}
\affiliation {Department of Physics, Applied Physics, and Stanford
Synchrotron Radiation Laboratory, Stanford University, Stanford,
CA 94305} \affiliation {Advanced Light Source, Lawrence Berkeley
National Lab, Berkeley, CA 94720}

\author{T. Yoshida}
\affiliation {Department of Physics, Applied Physics, and Stanford
Synchrotron Radiation Laboratory, Stanford University, Stanford,
CA 94305} \affiliation {Department of Complexity Science and
Engineering, University of Tokyo, Kashiwa, Chiba, Japan}

\author{D. H. Lu}
\affiliation {Department of Physics, Applied Physics, and Stanford
Synchrotron Radiation Laboratory, Stanford University, Stanford,
CA 94305}

\author{Y.L. Chen}
\affiliation {Department of Physics, Applied Physics, and Stanford
Synchrotron Radiation Laboratory, Stanford University, Stanford,
CA 94305}

\author{R.H. He}
\affiliation {Department of Physics, Applied Physics, and Stanford
Synchrotron Radiation Laboratory, Stanford University, Stanford,
CA 94305}

\author{Hsin Lin}
\affiliation {Physics Department, Northeastern University, Boston,
MA 02115}

\author{S. Komiya}
\affiliation {Central Research Institute of Electric Power
Industry, Iwato-kita, Komae, Tokyo, Japan}

\author{Y. Ando}
\affiliation {Central Research Institute of Electric Power
Industry, Iwato-kita, Komae, Tokyo, Japan}

\author{F. Zhou}
\affiliation {National Lab for Superconductivity, Institute of
Physics, Chinese Academy of Sciences, Beijing, China}

\author{W.X. Ti}
\affiliation {National Lab for Superconductivity, Institute of
Physics, Chinese Academy of Sciences, Beijing, China}

\author{J.W. Xiong}
\affiliation {National Lab for Superconductivity, Institute of
Physics, Chinese Academy of Sciences, Beijing, China}

\author{Z. X. Zhao}
\affiliation {National Lab for Superconductivity, Institute of
Physics, Chinese Academy of Sciences, Beijing, China}

\author{T. Sasagawa}
\affiliation {Department of Physics, Applied Physics, and Stanford
Synchrotron Radiation Laboratory, Stanford University, Stanford,
CA 94305} \affiliation {CREST, Department of Applied Physics,
Advanced Materials Science, and Superconductivity, University of
Tokyo, Bunkyo-ku, Tokyo, Japan}

\author{T. Kakeshita}
\affiliation {CREST, Department of Applied Physics, Advanced
Materials Science, and Superconductivity, University of Tokyo,
Bunkyo-ku, Tokyo, Japan}

\author{K. Fujita}
\affiliation {CREST, Department of Applied Physics, Advanced
Materials Science, and Superconductivity, University of Tokyo,
Bunkyo-ku, Tokyo, Japan}

\author{S. Uchida}
\affiliation {CREST, Department of Applied Physics, Advanced
Materials Science, and Superconductivity, University of Tokyo,
Bunkyo-ku, Tokyo, Japan}

\author{H. Eisaki}
\affiliation{Nanoelectronic Research Institute, AIST, Tsukuba
305-0032, Japan}

\author{A. Fujimori}
\affiliation {Department of Complexity Science and Engineering,
University of Tokyo, Kashiwa, Chiba, Japan}

\author{Z. Hussain}
\affiliation {Advanced Light Source, Lawrence Berkeley National
Lab, Berkeley, CA 94720}

\author{R. S. Markiewicz}
\affiliation {Physics Department, Northeastern University, Boston,
MA 02115}

\author{A. Bansil}
\affiliation {Physics Department, Northeastern University, Boston,
MA 02115}

\author{N. Nagaosa}
\affiliation {CREST, Department of Applied Physics, Advanced
Materials Science, and Superconductivity, University of Tokyo,
Bunkyo-ku, Tokyo, Japan}

\author{J. Zaanen}
\affiliation {The Instituut-Lorentz for Therorectical Physics,
Leiden University, Leiden, The Netherlands}

\author{T.P. Devereaux}
\affiliation {Department of Physics, University of Waterloo,
Waterloo, Ontario, Canada N2L 3G1}

\author{Z.-X. Shen}
\email[]{zxshen@stanford.edu} \affiliation {Department of Physics,
Applied Physics, and Stanford Synchrotron Radiation Laboratory,
Stanford University, Stanford, CA 94305} \affiliation {Advanced
Light Source, Lawrence Berkeley National Lab, Berkeley, CA 94720}

\date{November 30, 2006}

\begin{abstract}
To date, angle-resolved photoemission spectroscopy has been
successful in identifying energy scales of the many-body
interactions in correlated materials, focused on binding energies
of up to a few hundred meV below the Fermi energy. Here, at higher
energy scale, we present improved experimental data from four
families of high-$T_c$ superconductors over a wide doping range
that reveal a hierarchy of many-body interaction scales focused
on: the low energy anomaly ("kink") of 0.03-0.09eV, a high energy
anomaly of 0.3-0.5eV, and an anomalous enhancement of the width of
the LDA-based CuO$_2$ band extending to energies of $\approx$ 2
eV. Besides their universal behavior over the families, we find
that all of these three dispersion anomalies also show clear
doping dependence over the doping range presented.
\end{abstract}

\maketitle

\section{\label{introp}Introduction}

Many-body interaction is a key to understanding novel properties
of quantum matter.  As an extreme example, the complexity due to
charge, spin, and lattice interactions in high-$T_c$
superconductors makes it difficult to identify the essential
microscopic ingredients for the basic model - a reason behind the
current debate on the mechanism. The energy-momentum dispersion
relationship measured by angle-resolved photoemission spectroscopy
(ARPES) provides an excellent tool for identifying these scales.
Energy scales where these interactions are manifest usually
provide important insights into the nature of the interactions. At
an early stage, APRES proved to be successful in identifying the
energy scale of the d-wave gap in high-T$_c$ superconductors
\cite{gap:ZX, Group:Review}. To date, a focus of the discussion in
ARPES has been on the nature of electron-boson coupling, which
manifests itself in the form of the low energy anomaly ("kink")
near 0.03-0.09eV
\cite{Group:Review,Mode:Pasha,Mode:Kaminski,Mode:Johnson,Mode:Lanzara,Universal:Xingjiang,B1gMode:Tanja,MultiMode:XJZhou,CaxisScreening:Non}.
However, little attention has been paid to the features at higher
binding energies.

Given the strong many-body interactions and complex band
structure, one expects only incoherent features and complex
spectral weight modulation at high energy. The fact that one can
see neatly defined momentum dependent features which are robust
against doping and measuring conditions (e.g. photon energy) is
unexpected and thus provides a new opportunity to understand
many-body effects beyond what have traditionally been the points
of focus by ARPES, other than some early work on oxygen $p$ bands
\cite{Oband:Schabel,Oband:Hayn}. Here, we report ARPES experiments
on four families of high-$T_c$ cuprates over a wide range of
dopings: Bi$_{2}$Sr$_{2}$CuO$_{6}$ (Bi2201),
Bi$_{2}$Sr$_{2}$CaCu$_2$O$_{8}$ (Bi2212),
Ba$_{2}$Ca$_{3}$Cu$_4$O$_{8}$(O$_{\delta}$,F$_{1-\delta}$)$_2$
(F0234), and La$_{2-x}$Sr$_x$CuO$_4$ (LSCO). All of their measured
energy-momentum dispersion relationships reveal the simultaneous
presence of three energy scales, marked as 1-3 in Fig. 1d: 1) the
band bottom, at nearly 2 eV in optimally-doped Bi2201, which is
deeper than predicted by LDA calculation, 2) the high-energy
anomaly (HEA) at $\approx$ 0.3-0.5 eV (green arrow) and 3) the
low-energy "kink" (LEK) near 70 meV (red arrow) which can be
better seen in an enlarged energy window. Various aspects of these
energy scales, including the peculiar doping dependent effects,
will be discussed to obtain insights on many-body interactions and
their interplay in cuprates.

\begin{figure*}
\begin{center}
\includegraphics [width=6in, clip]{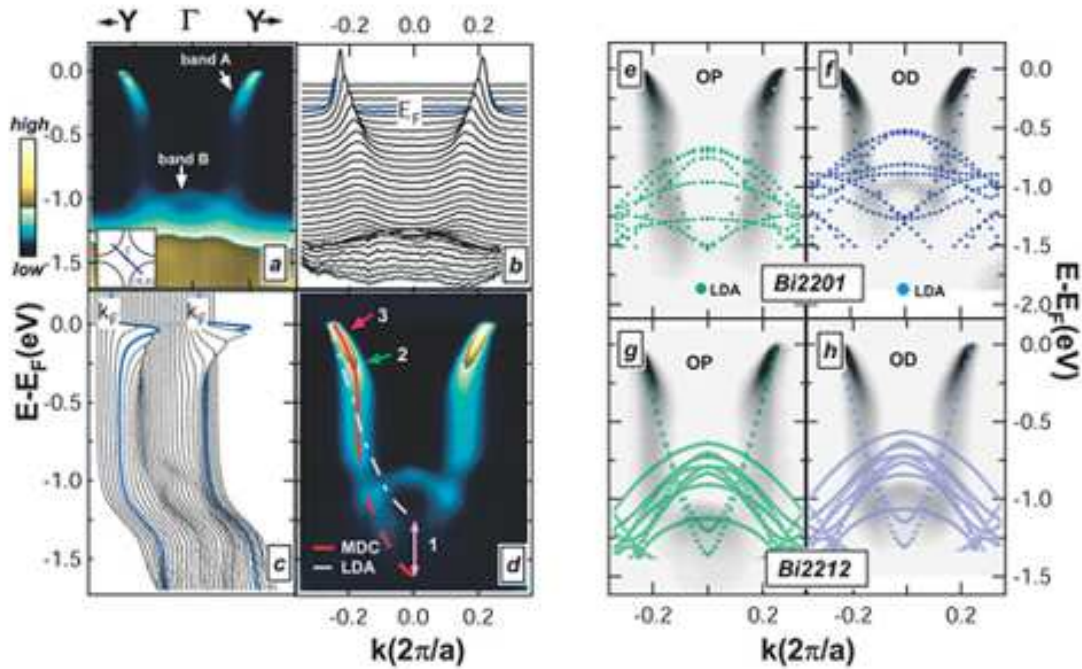}
\end{center}
\caption{\label{fig1} (a) Raw ARPES spectrum, along (0, 0) to
($\pi$, $\pi$) direction, of non-superconducting overdoped Bi2201
at T = 30 K while (b) and (c) represent its MDCs and EDCs
respectively. (d) the raw data is normalized with angle-integrated
EDC profile for clearer view and the numbers 1-3 mark the wider
band width, the high and low-energy anomalies respectively.
Normalized ARPES spectra are compared with LDA calculations as
follows: (e) OP Bi2201 with T$_c$ = 35K (T=45K), (f) OD Bi2201,
non-superconducting (T=30K), (g) OP Bi2212 with T$_c$ = 96K (T =
110K, LDA from Ref. \cite{LDA:Bansil}) and (h) OD Bi2212 with
T$_c$ = 65K (T=76K). Inset in Fig. 1a shows the momentum space of
the data. Note that the LDA bands in panel (h) are obtained by
rigidly shifting the bands in (g) to account for the correct
doping level. It is not clear if feature B should be matched to
the top of the band structure in (e) or (f) at $\Gamma$; however,
this uncertainty is not important for our argument which only
requires a relative shift in going from the OP to the OD case.}
\end{figure*}

\section{\label{experiement}EXPERIMENT AND LDA CALCULATION}

We have measured four families of high-$T_c$ cuprates:
Bi$_{2}$Sr$_{2}$CuO$_{6}$ (Bi2201),
Bi$_{2}$Sr$_{2}$CaCu$_2$O$_{8}$ (Bi2212),
Ba$_{2}$Ca$_{3}$Cu$_4$O$_{8}$(O$_{\delta}$,F$_{1-\delta}$)$_2$
(F0234) \cite{F0234:Yulin}, and La$_{2-x}$Sr$_x$CuO$_4$ (LSCO).
The Bi2201 samples are optimally-doped of Tc = 35K and
non-superconducting overdoped. The Bi2212 samples are
optimally-doped of Tc = 92K and overdoped of Tc = 65K. The F0234
samples are of Tc = 60K. And, the LSCO samples has a wide range of
dopings: x = 0.03, 0.05, 0.063, 0.07, 0.075, 0.09, 0.12, 0.15,
0.22 and 0.3. The measurements were carried out on beamline 10.0.1
at the ALS, using a Scienta R4000 electron energy analyzer. This
analyzer has the advantage of a large-angle window which can cover
the band dispersion across the Brillouin zone as shown in Fig. 1.
We stress that the wide angle scan allowed us to record above data
without resorting to manual symmetrization. The photon energies
are 37, 40, 41,...,45 and 55 eV. The energy resolution between 12
and 20 meV was used for various measurements on different samples,
and the angular resolution is 0.3 degree. The samples were cleaved
in situ in vacuum with a base pressure better than $4 \times
10^{-11}$ torr. The samples were measured both in normal and
superconducting states.

LDA results here are based on full-potential well-converged
computations for the appropriate lattice structures, described in
greater details in Ref.\cite{LDA:Bansil} and \cite{LDA:Seppo}.

\section{\label{results}RESULTS}

Fig. 1a shows the raw ARPES image of the strongly overdoped Bi2201
sample while its raw momentum-distribution-curves (MDC) and
energy-distribution-curves (EDC) are shown in Fig. 1b and c,
respectively. To see the band near the bottom more clearly, the
raw ARPES image is divided by its profile of angle-integrated EDC
\cite{edcp:osterwalder}, as shown in Fig. 1d; this procedure will
not change the MDC-peak position at any given energy. We also note
that this renormalization procedure which is used in Fig. 1d-h and
Fig.2 is only for the purpose of displaying the lower and higher
energy features together since otherwise the intensity at the
higher-energy region will be too high to have a reasonable
displaying contrast. The raw data without this renormalization
procedure can be seen in Fig. 1a-c and Fig. 6a-i.

As marked by numbers 1-3 in Fig. 1d, we will focus on how our new
data reveal the simultaneous presence of three energy scales as
follows.

\subsection{\label{lda}anomalous
enhancement of the LDA-based CuO$_2$ band width}

\begin{figure} [t]
\includegraphics [width=3.5in, clip]{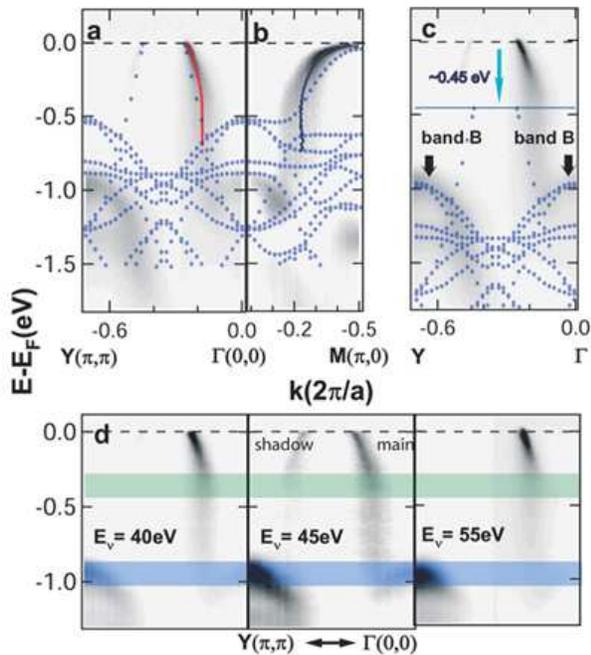}
\caption{\label{fig2} The comparison of normalized ARPES spectra
and LDA calculations along (a) ($\pi$, $\pi$) to (0,0) or nodal
direction and (b) (0,0) to ($\pi$, 0) or antinodal direction in OD
non-superconducting Bi2201 system. (c) shows the agreement of
ARPES spectrum and LDA calculation of high-energy band when
shifting the LDA by 450 meV to the higher energy. (d) The
comparison of the spectra at photon energies, E$_{\nu}$ = 40, 45
and 55 eV where the green shaded region denotes the energy scale
of HEA and blue shaded region indicates the top of the band B. We
note that for E$_{\nu}$ = 55 eV, due to the matrix element effect,
the intensity of the left band B is large and therefore we have
adjusted the color scale so that we can see the energy scales of
both band A and B clearly.}
\end{figure}

As shown in Fig. 1e-1h, ARPES spectra of optimally-doped (OP) and
overdoped (OD) samples of Bi2201 and Bi2212 systems are overlaid
on the corresponding LDA calculations. As seen in all measured
samples, the first peculiar feature, especially at lower doping,
is that the ARPES band width is found to be wider than LDA
calculation. This is anomalous as one expects interactions to
enhance the mass and reduce the band width. By extrapolating the
band (e.g. the red dashed line in Fig. 1d), one can get an
estimate of the band width that is suitable for qualitative
discussion. With doping, the discrepancy between the band widths
obtained from ARPES and LDA seems to be reduced.

Additional evidence that these high energy dispersions still
contain useful information comes from band B in Fig. 1a which
shows a maximum at $\Gamma$ point near 1 eV. This band has a
correspondence to an LDA band and thus provides confidence in the
data at higher energy scales, which has been largely unexplored in
the cuprates. From the LDA calculation with orthorhombic
distortion, this band is the band at Y point (left arrow, Fig.
2c), which is folded around the $(\pi/2,\pi/2)$ point. We then
compare the LDA and ARPES top part of this band B at $\Gamma$
point by shifting the LDA band down. A good agreement of ARPES and
LDA of this concave-down band (see Fig. 2c) can be obtained if the
LDA is shifted down $\approx$ 0.45 eV for OD Bi2201 sample and the
shifted energy increases to $\approx$ 0.8 eV for OP Bi2201
sample\cite{LDA:others}, leading to a filled band width near 2 eV.
A similar behavior is also observed in the Bi2212 system
\cite{LDA:others} (Fig. 1g-h). This band width enhancement was
also seen earlier in undoped Ca$_{2}$CuO$_2$Cl$_2$ (CCOC) as its
high-energy dispersion matches with the LDA calculation shifted by
~0.7 eV \cite{CCOC:Filip}.

\begin{figure}
\begin{center}
\includegraphics [width=2.5in, clip]{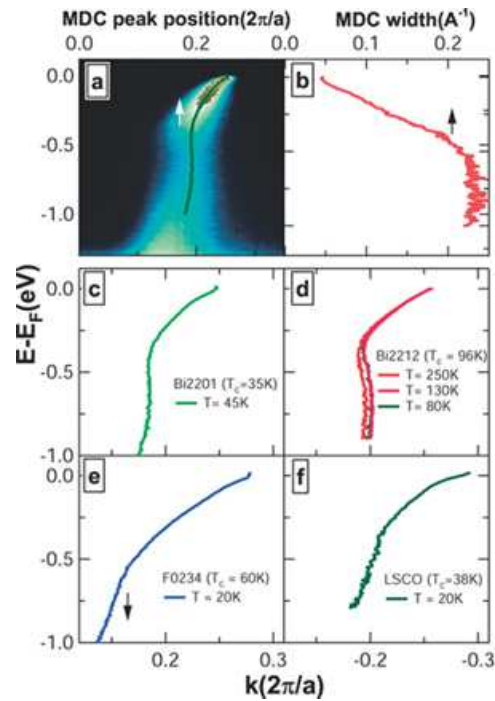}
\end{center}
\caption{\label{fig3:universal} (a) shows MDC-peak dispersion
plotted on top of ARPES spectrum and (b) shows corresponding MDC
width in OP Bi2201 system (T$_c$ = 35K). (c)-(f) show the
MDC-derived dispersions of Bi2201 (Tc = 35K), Bi2212 (Tc=65K),
LSCO(Tc=38K) and F0234 (Tc=60K) respectively. Additionally, the
temperature dependence of Bi2212 and LSCO dispersion is shown in
(d) and (f).}
\end{figure}

\begin{figure*}
\includegraphics [width=5.5in, clip]{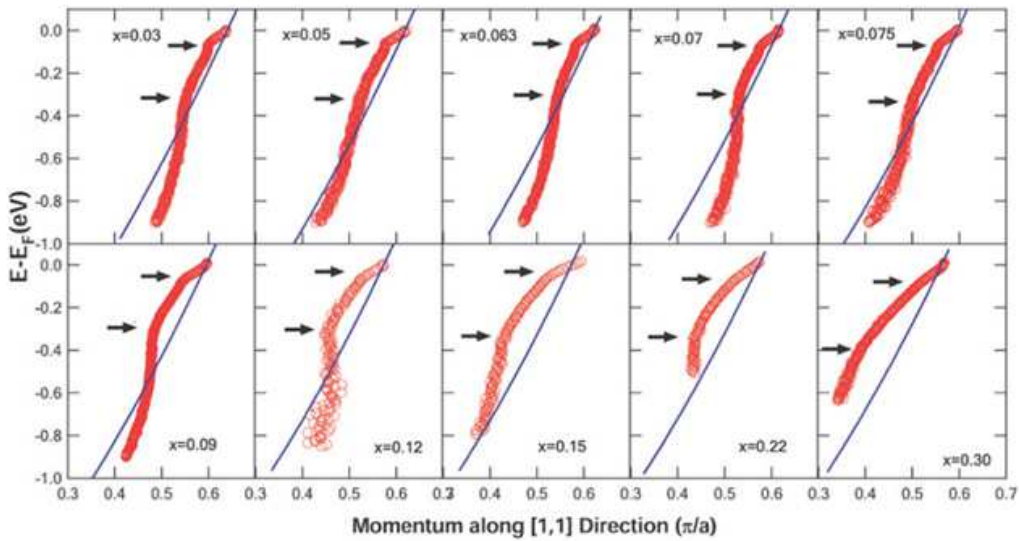}
\caption{\label{fig3} The doping dependence of high-energy
dispersion (red circles) in LSCO system at T = 20K at doping x =
0.03 to 0.30. LDA curves for various dopings (blue lines) are
obtained by appropriate rigid shifts of the computations for x=0.}
\end{figure*}

\subsection{\label{HEA}high energy anomaly of 0.3-0.5 eV}

Next, we discuss the high energy anomaly (HEA) near 0.3-0.5 eV. We
extract the MDC peak position by fitting to Lorentzian curves, as
shown by the red and blue curves in Fig. 2a and 2b. We note that
especially at high-binding energy, MDC peak position may not
represent the real dispersion. Explained in appendix, a full 2D
analysis, which directly extracts the spectral function, $A(k,
\omega)$ and matrix element term separately at once, can avoid the
problem of the MDC or EDC analysis alone. However, since the
kink-like structure is so large, the MDC-derived dispersion should
be able to approximately identify the energy scale of the HEA. To
check the 3 dimensional behavior of the band A and B, we perform
the measurements at various photon energies 40, 41, ..., 45 eV and
55 eV, probing different perpendicular momenta $k_z$. As shown in
Figs. 2d, we find that HEA scale and the top of Band B are not
very sensitive to the photo energies while as well these energy
scales in our LDA calculations do not show strong $k_z$
dependence.

This HEA is present in various cuprate families. While earlier
seen in undoped CCOC \cite{CCOC:Filip}, the MDC-peak-position
graphs of Bi2201, Bi2212, LSCO and F0234 plotted in Fig. 3, reveal
its universality. The energy scales are around 0.3-0.4 eV in
Bi2201, Bi2212 and LSCO while around 0.5-0.6 eV in F0234 (for
electron-doped band). HEA persists in both superconducting and
non-superconducting samples, albeit its strength depends on
doping. Fig. 4 shows the plots of MDC-peak position of LSCO
samples which cover a wide doping range, x = 0.03 to 0.30. From
the figure, the HEA energy does not change much with doping.
However, if we define the size of HEA to be the difference between
the MDC-derived and LDA dispersion, it increases upon doping in
this range. Similar doping-dependent behavior is also observed in
Bi2201 and Bi2212 samples which cover a narrower range of doping.
For superconducting samples, the HEA persists above and below Tc.
consistent with ARPES data, from in-plane optical conductivity of
Bi2212 \cite{InplaneOptical:vanderMarel}, Norman and Chubukov
recently report that the real part of the self-energy is large
with a maximum value around 0.3-0.4 eV
\cite{InplaneOptical:Chubukov}.

\subsection{\label{LEK}low energy "kink" of 0.03-0.09eV}

Finally, the low energy "kink" (LEK) around 0.03-0.09 eV is
indicated with arrows in Fig. 2d, and upper arrows in Fig. 4 and
Fig. 5a. Since this feature has been already discussed with regard
to the interaction of electron to sharp bosonic mode(s)
\cite{Group:Review,Mode:Pasha,Mode:Kaminski,Mode:Johnson,Mode:Lanzara,Universal:Xingjiang,B1gMode:Tanja,MultiMode:XJZhou,CaxisScreening:Non},
we will not go into the details of this feature, except commenting
upon its interesting doping dependence. For LEK, the size of this
feature, which is interpreted as strength of electron-boson
coupling, reduces upon doping in LSCO \cite{MultiMode:XJZhou} and
Bi2201 \cite{CaxisScreening:Non} while the size of HEA defined
previously increases upon doping. It is then intriguing to ask
whether an interplay of these two scales of low and high-energy
anomalies will affect our understanding of the doping dependent
effects seen in cuprates.

\section{\label{discussion}DISCUSSION}

\begin{figure}
\begin{center}
\includegraphics [width=3.6in, clip]{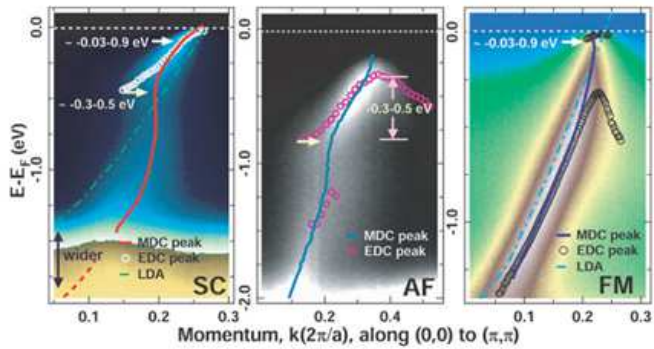}
\end{center}
\caption{\label{fig4} Comparison of electronic structures of three
perovskites in (a) superconducting (SC), (b) antiferromagnetic
(AF) and (c) ferromagnetic (FM) phases. The images show ARPES data
of (a) optimally-doped Bi2201, (b) undoped CCOC, and (c) LSMO.}
\end{figure}

The presence of three energy scales in the same data set hints at
the hierarchy of interactions that are important to the dynamics
of electrons in cuprates. Aside from LEK which we believe is
caused by electron-phonon interaction
\cite{Mode:Lanzara,B1gMode:Tanja,MultiMode:XJZhou,CaxisScreening:Non},
HEA and the expanded band width are new observations that require
more discussion.

To gain more insights into the nature of the energy scales
observed, Fig. 5 compares the data of the superconducting (SC)
sample, OP Bi2201, with that of the antiferromagnetic (AF) parent
compound CCOC and ferromagnetic (FM)
La$_{1.2}$Sr$_{1.8}$Mn$_2$O$_7$ (LSMO). Comparison of LSMO and
cuprates may give us some insight since LSMO is reported to have
similar pseudogap behavior \cite{LSMO:Norman}. The LSMO comparison
yields two insights: i) HEA is probably related to
antiferromagnetism as it is absent in the ferromagnetic state and
ii) the LDA calculation apears to give a correct band width if
antiferromagnetism is not present. As for LEK near 0.03-0.09 eV,
it is seen in metallic cuprates and LSMO but not in insulating
CCOC where the polaron effect is too strong, thus suppressing the
quasiparticle weight dramatically \cite{Polaron:Kyle}.

HEA may be related to the short range Coulomb interaction as in
the Hubbard (or t-J) model. Calculations using these models show
that in undoped Sr$_{2}$CuO$_2$Cl$_2$, the quasiparticle band
width of the part below HEA is set by the J scale to be around 2-3
J $\simeq$ 0.25-0.35 eV \cite{tJ:Dagotto} while the higher energy
part is presumed to be the incoherent band with t scale
\cite{CCOC:Filip}. HEA in the Hubbard calculation, which may come
from the meeting of the quasi-particle band and the incoherent
lower-Hubbard band, can be seen from the slightly underdoped to
overdoped regime \cite{Hubbard:Alexandru}; however, a further
check still remains in the small doping regime where the
calculation is challenging. In SrVO$_3$, a feature similar to HEA
is seen in ARPES \cite{SrVO3:Teppei} and in LDA+DMFT calculation
\cite{SrVO3:Anderson}. With these models, the size of HEA is
expected to be less pronounced upon doping \cite{lowerhubbard:Tom}
which is opposite to the doping effect we see in Fig. 4.

One possible way to understand the anomalous doping dependence
could be the interplay between electron-electron and
electron-phonon interactions.  This drives the system into the
polaronic regime in the underdoped samples \cite{Polaron:Kyle},
making a quantitative analysis of the size of the HEA difficult
because obtaining polaronic physics correctly via an MDC analysis
is challenging. Since the polaron physics is very strong in
lightly-doped regime, this may artificially suppress the HEA size.

Another possibility is that HEA comes from the renormalization
effect by bosonic mode(s). The in-plane plasmon and the two-magnon
mode are two possible candidates. The coupling strength of
in-plane plasmon increases upon doping, as shown by the
energy-loss function in Ref. \cite{abplasmon:Uchida}.
Qualitatively, the in-plane plasmon should give a very similar
doping effect to that shown in Fig. 4. The problem is that the
plasmon mode energy in LSCO is around 0.8 eV, or twice as large.
On the other hand, the two-magnon mode energy is right in the
window for the hole-doped cuprates (~0.3-0.4eV) but not for FO234,
electron-doped band (~0.5-0.6 eV) while its strength quickly
reduces upon doping \cite{magnon:Sugai}.

Given the above alternative interpretations and issues related to
them, despite the quantitative problem of the HEA size as a
function of doping, the anomalous energy scale is still likely
caused by the Mott-Hubbard physics. The fact that HEA is absent in
FM LSMO (Fig. 5) but present in the Mott-Hubbard system SrVO$_3$
suggests that it is related to antiferromagnetism and Mott-Hubbard
physics.

Next is the enhancement of LDA band width which is anomalous
because interactions usually reduce the band width. One possible
candidate here could be the poor screening effect. With the
Hartree-Fock equation of free electron, the Coulomb exchange term
without the screening effect \cite{poorscreening:ashcroft} will
give rise to a negative term which results in an increased band
width; the effect of poor screening can be seen in semiconductors.
In cuprates, the electrons in lightly-doped systems are poorly
screened compared to the overdoped regime and hence the
discrepancy of band width is larger in underdoped systems. We note
here that, for the SrVO$_3$ system, LDA+DMFT calculation
\cite{SrVO3:Anderson} gives the band width correctly. This
distinction between SrVO$_3$ and cuprates should be investigated
further.

\section{\label{conclusion}CONCLUSION}

In conclusion, based on data from four families of cuprates over a
wide doping range, we present evidence of a hierarchy of multiple
energy scales in cuprates focused on: the low energy anomaly of
0.03-0.09 eV, a high energy anomaly of 0.3-0.5 eV, and an
anomalous enhancement of LDA band width, extending over an energy
scale of $\approx$ 2 eV. These results suggest that
electron-phonon interaction, short-range Coulomb interaction and
poor screening should be all considered to understand the nature
of cuprates.

\begin{acknowledgments}
We thank A. Macridin, B. Moritz, M. Jarrell, K.M. Shen, D.
Scalapino, D. van der Marel, N.J.C Ingle, F. Baumberger for
helpful discussions. The work at SSRL and ALS are supported by
DOE's Office of Basic Energy Sciences under Contracts No.
DE-AC02-76SF00515 and DE-AC03-76SF00098. The work at Stanford is
supported by NSF DMR-0304981 and ONR N00014-04-1-0048. W.M.
acknowledges DPST for the financial support. The work at
Northeastern is supported by the US DOE contract DE-AC03-76SF00098
and benefited from the allocation of supercomputer time at NERSC
and Northeastern's Advanced Scientific Computation Center (ASCC).
The work at IOP-CAS is supported by NSF of China. T.P.D.
acknowledges ONR N00014-05-1-0127.
\end{acknowledgments}

\appendix*
\section{a full 2D analysis}

In the following, we will show that interestingly, a 2D analysis
could describe ARPES data well, covering a large energy and
momentum space, by a compact set of parameters. This is an attempt
to go beyond the conventional EDC or MDC analysis alone. However,
we note that the physical meaning of such a parameterization still
remains to be explored.

\begin{figure*}
\begin{center}
\includegraphics [width=5.5in, clip]{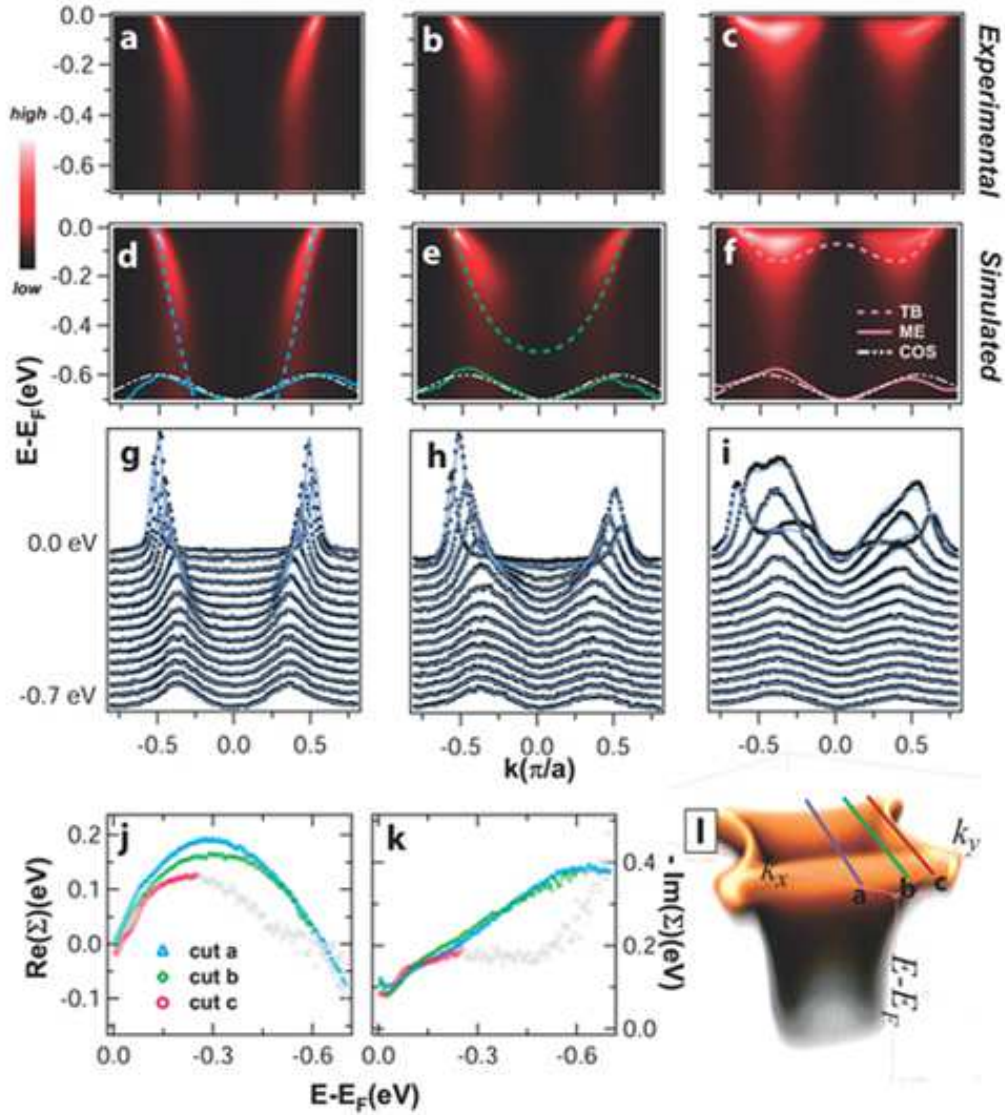}
\end{center}
\caption{\label{fig6} On the first row, a)-c) are the experimental
ARPES data along momentum direction as indicated by the band shown
in bottom right. On the second row, d)-f) are the corresponding
simulated images from the 2D analysis. The thicker color dashed
lines are the dispersions generated from the tight-binding (TB)
parameters given above. The solid color lines are the matrix
elements (ME) obtained from the 2D analysis where the smaller
white dashed lines are the empirically guessed form of the matrix
element in the form of a cosine function (COS). On the third row,
g)-i) are the MDCs of raw data (black dots) and corresponding
simulated images (blue line). j) and k) are the extracted real and
imaginary parts of the so-called self-energy in Eq. A2 for the
cuts a, b and c; the extracted values are plotted in colors up to
the energy not far from the bottom of the bare band (up to 0.6 eV
for cut b and 0.25 eV for cut c) and in grey at higher energy. l)
shows the band structure generated from the TB parameters and the
loosely-called self-energy along the nodal direction.}
\end{figure*}

Here we will apply the 2D analysis on the ARPES data of
Pb-substituted Bi2201. The overdoped (OD) samples,
Pb$_{0.38}$Bi$_{1.74}$Sr$_{1.88}$CuO$_{6+\delta}$, are
non-superconducting ($\rm T_c$ $<$ 4 K). ARPES data were collected
with a photon energy of 42 eV. The energy resolution was set to 18
meV. The linear polarization of the light source is fixed to be
in-plane along (0,0) to ($\pi$,$\pi$) throughout the measurement.
Note that the fitted matrix element, which is shown in Fig. 6d-i,
refers to this particular experimental geometry.

The intensity measured in an ARPES experiment on a 2D material
here will be parameterized by \cite{Group:Review}
\begin{equation}
I({\bf k},\omega) =I_0({\bf k},\nu,{\bf A}) f(\omega){\mathcal
A}({\bf k},\omega)
\end{equation}
where $I_0({\bf k},\nu,{\bf A})$ is proportional to the
one-electron matrix element and dependent on the polarization,
momentum and energy of the incoming photon, $f(\omega)$ is the
Fermi function. $\mathcal A({\bf k},\omega)$ is the
single-particle spectral function given by
\begin{equation}
{\mathcal A}({\bf k},\omega)  = \frac{(-1/\pi)\ \ \textrm{Im}
\Sigma({\bf k},\omega)}{[\omega -
\epsilon^{0}_{k}-\textrm{Re}\Sigma({\bf
k},\omega)]^{2}+[\textrm{Im}\Sigma({\bf k},\omega)]^{2}}
\end{equation}
where $\epsilon^{0}_{k}$ is the bare band dispersion, and
$\Sigma({\bf k},\omega)$ is the self-energy.

We note that we neglect the instrumental resolution here since the
feature of interest is large compared to the resolution. In the
following, we will assume weak momentum dependence of this
extracted self-energy (i.e. $\Sigma({\bf k},\omega)$ $\rightarrow$
$\Sigma (\omega)$ .)

Fig. 6 shows the comparison of the raw ARPES data (first row, Fig.
6a-c) and the parameterized data (second row, Fig. 6d-e). We
parameterize the ARPES data with the form given by Eq. A1 where
the spectral function ${\mathcal A}(k,\omega)$ is given by Eq. A2.
In the fitting, we do not assume any form of the self-energy and
the matrix element (i.e. every point of the fitted self-energy or
matrix element is a free parameter in the fitting procedure.) The
bare dispersion used here is the simple form given by the
tight-binding (TB) parameters which is fitted to the LDA
calculation shown in Fig. 1f and 2. The bare dispersion is given
by $E(k) = -2t[cos(k_x a)+ cos(k_y b)]-2 t' cos(k_x a) cos(k_y b)
- 2t''[cos(2 k_x a)+ cos(2 k_y b)] - E_F$ where the TB parameters
are $t = 0.435, t'= -0.1, t''= 0.038,$ and $E_F = -0.5231$ eV. For
simplicity, we use constant background in the fitting procedure.

The fitting of the ARPES data (Fig. 6d-e) is very well in
agreement with average error $ < 4\%$. Since the extracted
self-energy does not show a strong momentum dependence, the
approximate 3-dimensional spectra may be generated from this
information as shown in Fig. 6i.

\end{document}